%% file: ANA-EXOT-2018-48-PAPER.tex
\pdfoutput=1
\documentclass[cernpreprint,atlasdraft=true,texlive=2016,UKenglish,texmf]{atlasdoc}
 
\usepackage{atlaspackage}
\usepackage{atlasbiblatex}
\usepackage{atlasphysics}
\usepackage{colortbl}
\usepackage{multirow}
\usepackage{threeparttable}
\usepackage{cancel}
\usepackage{hhline}
 
\graphicspath{{logos/}{figures/}}
 
\usepackage{ANA-EXOT-2018-48-PAPER-defs}
 
\usepackage{comment}
\usepackage{bm}
 
\AtlasTitle{Search for $t\bar{t}$ resonances in fully hadronic final states in $pp$ collisions at $\sqrt{s}$ =13 TeV with the ATLAS detector}
\AtlasJournal{JHEP}
\AtlasJournalRef{\JHEP 10 (2020) 61}
\AtlasDOI{10.1007/JHEP10(2020)061}
\AtlasAbstract{
This paper presents a search for new heavy particles decaying into a pair of top quarks using
139 $fb^{-1}$ of proton--proton collision data recorded at a centre-of-mass energy of $\sqrt{s}$ = 13 TeV
with the ATLAS detector at the Large Hadron Collider.
The search is performed using events consistent with pair production of high-transverse-momentum
top quarks and their subsequent decays into the fully hadronic final states.
The analysis is optimized for resonances decaying into a $t\bar{t}$ pair with mass above 1.4 TeV, exploiting a dedicated multivariate technique with jet substructure
to identify hadronically decaying top quarks using large-radius jets and
evaluating the background expectation from data.
No significant deviation from the background prediction is observed.
Limits are set on the production cross-section times branching fraction for the new $Z'$ boson
in a topcolor-assisted-technicolor model.
The $Z'$ boson masses below 3.9 and 4.7 TeV are excluded at 95\% confidence level
for the decay widths of 1\% and 3\%, respectively.
}
\PreprintIdNumber{CERN-EP-2020-055}
 
\author{The ATLAS Collaboration}
\AtlasRefCode{EXOT-2018-48}
\hypersetup{pdftitle={ATLAS document},pdfauthor={The ATLAS Collaboration}}
 
\begin{document}
 
\maketitle
 
\tableofcontents
 
\newpage
\section{Introduction}
Discovery of new phenomena beyond the Standard Model (SM) is of great importance for high-energy particle physics.
Such discovery has the potential to shed light on unexplained observations in nature,
e.g. the large difference between the scales of electroweak interactions at ${\cal O}(100)$~\GeV\ and gravity at
${\cal O}(10^{19})$~\GeV, the Planck scale.
The top quark, the heaviest elementary particle in the SM, could provide a window to this new physics
through its large coupling to the scalar sector with a Higgs field.
Resonant production of top and anti-top quarks (\ttbar),
if observed, strongly indicates the presence of new particles, such as those predicted by topcolor-assisted-technicolor (TC2)~\cite{Hill:1993hs,Hill:1994hp,Harris:2011ez}, the two-Higgs-doublet model (2HDM)~\cite{Branco:2011iw} and
Randall--Sundrum (RS) models of warped extra dimensions~\cite{Randall:1999ee, Lillie:2007yh}.
These new particles could appear in the mass spectrum of the \ttbar system (\mtt) as a localized deviation from the SM prediction.
This paper presents a search for \ttbar
resonances in the \TeV\ mass range with subsequent decay into a fully hadronic final state ($\ttbar \to W^+bW^-\bar{b}$ with $W \to q\bar{q'}$),
performed using \lumi of proton-proton ($pp$) collision data recorded in 2015--2018 at \sqs = 13~\TeV\ with the ATLAS detector at the Large Hadron Collider (LHC).
 
The fully hadronic final state benefits from the largest top-quark decay branching fraction.
However, it poses challenges in reconstructing the \ttbar system using hadronic jets, and in discriminating a
new physics signal from SM production of multijet processes which have a high production cross-section. Exploiting dedicated top-quark identification techniques, both the ATLAS~\cite{EXOT-2016-24} and CMS~\cite{CMS-B2G-17-017} experiments demonstrated that the search in the fully hadronic final states can have
comparable sensitivity at \mtt above $\sim1$~\TeV\ to the analysis in a lepton+jets final state
(where one of the $W$-bosons decays into
an electron or muon and a neutrino).
This paper extends this further by adopting an advanced top-quark identification method based on a deep
neural network. In this high-mass range, the top-quark decay products become close enough for
the hadronically decaying top quark to be reconstructed as a single large-radius jet with a characteristic internal substructure.
The search focuses on resonances with the intrinsic decay width comparable to or smaller than the detector resolution,
by looking for a localized excess over the smoothly falling mass spectra of the reconstructed SM \ttbar candidates. The
background spectrum is derived from data by fitting a smoothly falling function to the \mtt distributions.

Searches for heavy particles decaying into a \ttbar pair have been performed with the ATLAS and CMS experiments
using $pp$ collisions at \sqs = 7~\TeV~\cite{TOPQ-2011-23,TOPQ-2012-14,CMS-EXO-11-006,CMS-TOP-11-010,CMS-TOP-12-017},
8~\TeV~\cite{CMS-B2G-13-001,TOPQ-2012-18,CMS-B2G-13-008,EXOT-2016-04} and
13~\TeV~\cite{EXOT-2015-04, EXOT-2016-24, CMS-B2G-16-015, CMS-B2G-17-017}.
Both experiments performed the searches in
the fully hadronic and lepton+jets final states using 13~\TeV\ data collected in 2015--2016, while CMS included
the dileptonic final states as well. The spin-1 colour-singlet boson in a topcolor-assisted-technicolor model~\cite{Hill:1993hs,Hill:1994hp},
used in the previous \ttbar resonance search in the fully hadronic final state
in ATLAS~\cite{EXOT-2016-24}, continues to be used as a benchmark in the results presented in this paper. This leptophobic $Z'$ boson (denoted by \zprime), referred to as Model IV in Ref.~\cite{topcolor2}, is
mainly produced by $q\bar{q}$ annihilation and decays into first- and third-generation quarks.
The model parameters are chosen to maximize the branching fraction of the $\zprime\rightarrow\ttbar$ decay,
which reaches 33\%, and the intrinsic decay width of the \zprime boson divided by its mass $m$
is set to $\Gamma/m=1$\% or 3\%.\footnote{In the rest of this paper, the decay width of a resonance divided by the resonance mass is referred to as the width.}
The predicted $\zprime\rightarrow\ttbar$ production cross section is about 2.0 (7.6)~fb at the \zprime mass of 4~\TeV\ and
$\Gamma/m=1$\% (3\%).
Among several benchmark signal models, the ATLAS searches excluded masses below 3.1 and 3.0~\TeV\
in the fully hadronic and lepton+jets final states, respectively, for the new \zprime boson
with $\Gamma/m=1\%$~\cite{EXOT-2015-04, EXOT-2016-24}.
The CMS search excluded the \zprime boson with $\Gamma/m=1\%$ up to 3.8~\TeV\ using the combination of
all three final states including the dileptonic final state~\cite{CMS-B2G-17-017}.
 
\section{ATLAS detector}
\newcommand{\AtlasCoordFootnote}{
ATLAS uses a right-handed coordinate system with its origin at the nominal interaction point (IP)
in the centre of the detector and the $z$-axis along the beam pipe.
The $x$-axis points from the IP to the centre of the LHC ring,
and the $y$-axis points upwards.
Cylindrical coordinates $(r,\phi)$ are used in the transverse plane,
$\phi$ being the azimuthal angle around the $z$-axis.
The pseudorapidity is defined in terms of the polar angle $\theta$ as $\eta = -\ln \tan(\theta/2)$.
Angular distance is measured in units of $\Delta R \equiv \sqrt{(\Delta\eta)^{2} + (\Delta\phi)^{2}}$.}
 
The ATLAS experiment uses a multipurpose, forward--backward symmetric detector\footnote{\AtlasCoordFootnote}
with nearly $4\pi$ solid angle coverage, as described in Refs.~\cite{PERF-2007-01,ATLAS-TDR-19,Abbott_2018}.
It consists of an inner tracking detector (ID) surrounded by a thin superconducting solenoid, electromagnetic (EM) and hadronic calorimeters, and a muon spectrometer.
 
The ID consists of a silicon pixel tracker,
a silicon microstrip tracker (SCT) and a transition radiation tracker, all immersed in a \SI{2}{\tesla} axial magnetic field,
and provides charged-particle tracking in the range $|\eta|<2.5$.
The EM calorimeter is a lead/liquid-argon (LAr) sampling calorimeter, divided into
a barrel section covering the range $|\eta|<1.475$ and two endcap sections covering $1.375<|\eta|<3.2$.
In the region $|\eta|<1.8$, an additional
thin LAr presampler layer is used to correct for energy losses in the material upstream
of the calorimeters.
The hadronic calorimetry is provided by a steel/scintillator tile sampling calorimeter
in the central region ($|\eta|<1.7$) and by a copper/LAr calorimeter in the endcap regions ($1.5<|\eta|<3.2$).
The forward region ($3.1<|\eta|<4.9$) is instrumented
with copper/LAr and tungsten/LAr calorimeter modules optimized for electromagnetic and hadronic measurements,
respectively.
Surrounding the calorimeters is a muon spectrometer that consists of three air-core superconducting toroidal magnets
and tracking chambers,
providing precision tracking for muons with $|\eta|<2.7$ and trigger capability for $|\eta|<2.4$.
 
A  two-level trigger system is used to select events for offline analysis~\cite{TRIG-2016-01}.
Events are first selected by the level-1 trigger implemented with custom electronics, which uses a subset of the detector information
to reduce the event rate to approximately \SI{100}{\kilo\hertz}. This is followed by a software-based trigger that
reduces the accepted event rate to an average of  \SI{1}{\kilo\hertz} by refining the level-1 trigger selection.
 
 
\section{Data and simulation samples}
\label{sec:data_sim}
The search is performed using \lumi\ of data recorded from $pp$ collisions at \sqs = 13~\TeV.
The analysis uses data collected during stable beam conditions with the relevant detectors operational.
Events are required to have at least one $pp$ interaction vertex with two or more tracks with transverse momentum (\pt)
greater than 500~\MeV.
If more than one vertex is found in an event, the one with the largest $\sum \pt^2$ of associated tracks is chosen as the primary vertex.
Simulated signal and background event samples are used to optimize the event selection,
to validate the background estimation technique and
to perform hypothesis testing of the benchmark \zprime signal model.
 
The main backgrounds after applying the event selection criteria (Section~\ref{sec:analysis}) are expected to be composed of events from
SM \ttbar and multijet production processes. The total number of background events, including events from other minor backgrounds, is
estimated directly from data using functional fits to the \mtt spectra, as detailed in Section~\ref{sec:background}.
However, simulated samples of SM \ttbar and multijet events are used to establish the background estimation technique.
 
For SM $t\bar{t}$ production,
the next-to-leading-order (NLO) Monte Carlo (MC) generator \POWHEGBOXV{v2}~\cite{Nason:2004rx,Frixione:2007vw,Alioli:2010xd} was used
with the NNPDF3.0 NLO~\cite{Ball:2014uwa}
parton distribution function (PDF) set in the matrix element calculations.
The \ttbar production cross-section is scaled to a next-to-next-to-leading-order (NNLO) calculation in QCD
including resummation of
next-to-next-to-leading logarithmic soft gluon terms with Top++2.0~\cite{Cacciari:2011hy,Beneke:2011mq,Baernreuther:2012ws,Czakon:2012zr,Czakon:2012pz,Czakon:2013goa,Czakon:2011xx}.
Parton showering, hadronization and the underlying event were
simulated using \PYTHIAV{v8.230}~\cite{Sjostrand:2007gs} with the leading-order (LO)
NNPDF2.3~\cite{Ball:2012cx} PDF set
and the A14 set of tuned parameters~\cite{ATL-PHYS-PUB-2014-021}.
The $h_{\mathrm{damp}}$ parameter, which controls the transverse momentum of the first additional parton emission
beyond the Born level, was set equal to the top-quark mass~\cite{ATL-PHYS-PUB-2016-020}.
The top-quark kinematics in $t\bar{t}$ events were corrected to account for electroweak higher-order
effects~\cite{Kuhn:2013zoa}.
The generated events were weighted by this correction factor as a function of the flavour and centre-of-mass energy of the initial partons,
and of the decay angle of the top quarks in the centre-of-mass frame of the initial partons.
The value of the correction factor decreases with increasing \mtt from 0.98 at $\mtt=0.4$~\TeV\ to 0.87 at
$\mtt=3.5$~\TeV.
Multijet processes were simulated with \PYTHIAV{v8.186}~\cite{Sjostrand:2007gs}
using the NNPDF2.3 LO PDF set and the A14 set of tuned parameters for the underlying event.
 
Simulated signal samples of a spin-1 \zprime boson decaying into a \ttbar pair were generated using \PYTHIAV{v8.165}
with the NNPDF2.3 LO PDF set and the A14 set of tuned parameters. The production cross-section is scaled to a
NLO prediction by multiplying by a factor 1.3~\cite{Bonciani:2015hgv}
and the coupling parameters of the \zprime boson are chosen to be the same as those used in Ref.~\cite{EXOT-2016-24}.
The signal samples were generated for \zprime masses of {1750, 2000, 2250, 2500, 2750, 3000, 4000 and 5000}~\GeV.
A sum of Crystal Ball\footnote{A Crystal Ball function is composed of a Gaussian distribution at the core, connected with a power-law distribution describing the lower tail.} and Gaussian functions (CB + Gauss) is used to model the signal \mtt
distributions and create an interpolated signal \mtt template for masses where MC samples are not available. The fit function takes the form:
 
\begin{equation}\nonumber
J\left(x;f,\mu,\sigma, \alpha_\text{CB},n_\text{CB},\mu_\text{CB},\sigma_\text{CB}\right) = f\cdot g\left(x;\mu,\sigma\right) + \left(1-f\right) h\left(x;\alpha_\text{CB},n_\text{CB},\mu_\text{CB},\sigma_\text{CB}\right)
\end{equation}
 
where $g\left(x;\mu,\sigma\right)$ represents a Gaussian function with parameters $\mu$ and $\sigma$, $h\left(x;\alpha_\text{CB},n_\text{CB},\mu_\text{CB},\sigma_\text{CB}\right)$ represents a Crystal Ball function with parameters $\alpha_\text{CB}$, $n_\text{CB}$, $\mu_\text{CB}$ and $\sigma_\text{CB}$ and $0<f<1$ is a fractional coefficient. The parameters $\mu$ and $\sigma$ represent the mean value and the width of the Gaussian function, and
the $\alpha_\text{CB}$ and $n_\text{CB}$ are the threshold and exponent parameters of the CB function, respectively. Thus there are seven shape parameters and an additional normalization parameter. The parameters for the interpolated signal templates are obtained by using either a linear or polynomial interpolation between the values of parameters estimated using a CB + Gauss fit to signal MC \mtt distributions. The interpolated signal \mtt templates are used at masses of 1875, 2125, 2375, 2625, 2875, 3250, 3500, 3750, 4250, 4500 and 4750~\GeV.
The signal MC samples are also used to evaluate the acceptance and selection efficiencies for the signals considered
in the search.

The {\textsc{EvtGen}}~v1.2.0 program~\cite{Lange:2001uf} was used in all simulated samples to model the properties of
bottom and charm hadron decays.
All simulated samples include the effects of multiple \pp interactions in the same and neighbouring
bunch crossings (pile-up) and were processed through the ATLAS detector simulation~\cite{SOFT-2010-01}
based on $\GEANT4$~\cite{Agostinelli:2002hh}.
Pile-up effects were emulated by overlaying simulated minimum-bias events generated with \PYTHIAV{v8.186}
using the MSTW2008LO PDF set~\cite{Martin:2009iq} and the A2 set of tuned parameters~\cite{ATL-PHYS-PUB-2012-003}.
The number of overlaid minimum-bias events was adjusted to match the observed data.
Simulated events are processed through the same reconstruction software as the data, and corrections are applied so that the object identification efficiencies, energy scales and energy resolutions match those determined from control samples of data.
 
\section{Event reconstruction and selection}
\label{sec:analysis}
At high transverse momentum with \pt above approximately 350~\GeV, the decay products of
a hadronically decaying top quark can be reconstructed as a single large-radius ($R=1$) jet. Such large-$R$ jets are
characterized by the presence of multiple cores associated with the `subjets' from a $b$-hadron and a
$W \to q\bar{q'}$ decay.
One of the main backgrounds in this search, multijet production, is dominated by jets from light-flavour quarks
and gluons. Typically, such jets have single cores, but would have enhanced contributions of multiple cores associated with
large-angle emission when a conventional requirement of large jet mass is applied in the selection of top-quark candidate jets.
Therefore, to discriminate genuine top-quark jets from multijet background
with a mass around the top-quark mass, an improved analysis technique that takes advantage of more-detailed
jet substructure information is desired.
An additional challenge in high-mass \ttbar resonance searches comes from the fact that
the subjet containing a $b$-hadron ($b$-jet) in top-quark decays has, in close vicinity,
hadronic activity from $W \to q\bar{q'}$ decays. Therefore, the identification of a $b$-jet inside a large-$R$ jet
will need to be optimized to reduce contributions from non-$b$-hadron decays.
The analysis presented here addresses these challenges by adopting advanced top-quark tagging and $b$-jet identification techniques,
as detailed below.

\subsection{Object reconstruction}
Large-$R$ jets are built from three-dimensional topological clusters of energy deposits in the calorimeter, calibrated to
the hadronic energy scale with the local cluster weighting (LCW)~\cite{PERF-2014-07} procedure, using the anti-$k_t$ algorithm~\cite{Cacciari:2008gp}
with a radius parameter $R=1.0$.
The non-compensating calorimeter response and the energy loss in dead material or
due to out-of-cluster leakage of deposited energy are accounted for in the LCW procedure.
The reconstructed jets are `trimmed'~\cite{Krohn:2009th} to reduce contributions from pile-up and soft interactions.
This is performed by reclustering the jet constituents into subjets using
the $k_t$ algorithm~\cite{Catani:1991hj,Ellis:1993tq,Catani:1993hr} with a radius parameter $R=0.2$ and discarding
subjets with \pt less than 5\% of the \pT of the parent jet~\cite{ATLAS-CONF-2015-035}.
The four-momenta of large-$R$ jets are finally reconstructed from the momentum vectors of the remaining subjets and
corrected using simulation~\cite{PERF-2011-03,PERF-2012-01}.
The mass of the large-$R$ jet, \mJ, is calculated by combining the calorimeter
energy measurement with the track information from the ID to mitigate the effect on the mass resolution
from the limited angular granularity of the calorimeter, as in Ref.~\cite{ATLAS-CONF-2016-035}.
The large-$R$ jets considered in the analysis are selected by requiring
$\pt > 200$~\GeV, $|\eta| < 2.0$ and $\mJ>50$~\GeV.
 
The identification of hadronically decaying top quarks that are reconstructed using large-$R$ jets is performed using a multivariate classification algorithm employed in
a deep neutral network~\cite{JETM-2018-03}. In the kinematic regime of interest in this search, a single large-$R$ jet
captures the top-quark decay products, resulting in a characteristic multi-core structure within the jet, in contrast to a typical
single-core structure associated with jets in multijet background processes.
In order to exploit this characteristic behaviour for the top-quark identification,
a multivariate top-tagging classifier has been developed~\cite{JETM-2018-03}.
The tagger uses multiple jet-level discriminants as inputs,
e.g.\ calibrated jet \pT and mass, information about the dispersion of the jet constituents such as
$N$-subjettiness~\cite{Thaler:2010tr,Thaler:2011gf},
splitting scales~\cite{Thaler:2008ju} and energy correlation functions~\cite{Larkoski:2013eya,Larkoski:2014gra}.
 
The tagger used in this analysis is optimized for top-quark-initiated jets that satisfy the `contained' criteria
defined using the simulation as follows. First, particle-level large-$R$ jets are built from all stable particles (with $c\tau>10$~mm)
at the generator level using a radius parameter $R=1.0$ and trimmed in the same way as for reconstructed jets in data.
A trimmed particle-level jet is required to match a generator-level top quark within $\Delta R<0.75$,
have a mass above 140~GeV and at least one ghost-associated~\cite{Cacciari:2008gn} $b$-hadron;
hereafter the selected particle-level jet is referred to as a truth-contained jet.
A reconstructed detector-level large-$R$ jet is then required to satisfy the same $\Delta R$ requirement
with respect to the matched particle-level jet to ensure that the jet contains the top-quark decay products.
The top-tagger calibrated for such large-$R$ jets turns out to be less sensitive to generator differences
in the determination of top-quark decay products falling inside a jet,
compared with the definition used in Ref.~\cite{JETM-2018-03}.
In this analysis
the `top-tagged' large-$R$ jets are selected using requirements on the classifier corresponding to an efficiency of 80\%
over the range of generator-level top-quark \pt relevant for the considered \zprime signal.
With the requirement of 80\% efficiency,
the rejection factor for light-flavour quark and gluon jets is approximately 30 (12) at a top-quark \pt of 500 (3000)~\GeV.
 
Jets built from charged-particle tracks reconstructed in the ID (track-jets) are used to identify jets
containing $b$-hadrons.
The tracks are first selected by requiring them to be associated with the primary vertex and to contain
a minimum number of hits in the pixel and SCT detectors, and then requiring them to have
$\pt>500$~\MeV\ and $|\eta|<2.5$.
In the topology of highly boosted top quarks~\cite{JETM-2018-03}, the $b$-hadron decay product is surrounded by
other hadronic activity from $W \to q\bar{q'}$ decays and this additional contribution from nearby particles degrades
the identification of charged-particle tracks from the $b$-hadron decay.
To overcome this, the track-jets are built from the selected tracks using the anti-$k_t$ algorithm with
$R$ varying as a function of the jet \pt.
This `variable-radius' (VR) track-jet~\cite{Krohn:2009zg} has an effective jet radius, $R_{\text{eff}}$,
proportional to the inverse of the \pt of the jet in the jet-finding procedure:
\begin{equation}\nonumber
R \to R_{\text{eff}}(\pt)=\rho/\pt,
\end{equation}
where the $\rho$-parameter that controls the effective radius is set to $\rho=30$~\GeV. There are two additional parameters, $R_{\text{min}}$ and $R_{\text{max}}$, used to set the minimum and maximum bounds on the
jet radius, and these are set to 0.02 and 0.4, respectively~\cite{ATL-PHYS-PUB-2017-010}. The values of these parameters are
determined by examining the efficiency of identifying two $b$-jets within a large-radius jet associated with a high-\pt Higgs
boson decaying into a $b$-quark pair~\cite{ATL-PHYS-PUB-2017-010}.
 
The VR track-jets are composed of at least two constituent tracks and are required to
have $\pt>10$~\GeV\ and $|\eta|<2.5$.
The VR track-jets containing $b$-hadrons are identified using the `DL1'
algorithm~\cite{FTAG-2018-01}.
This algorithm is based on a multivariate classification technique with an artificial deep neural network to combine
information from the impact parameters of displaced tracks, reconstructed muons in jets,
and topological properties of secondary and tertiary decay vertices reconstructed within the jet.
The $b$-jets are selected in the analysis using the requirement corresponding to an efficiency of
77\% for identifying $b$-jets in simulated SM \ttbar events. This requirement has
corresponding rejection factors of 5 and 128 for jets containing $c$-hadrons and light-flavour jets,
respectively.
Efficiencies to identify $b$-jets, $c$-jets, and light-flavour jets are corrected in the simulation
to account for deviations from the efficiencies observed in data~\cite{FTAG-2018-01}.
 
Events containing charged leptons (electrons or muons) are removed in this analysis to avoid overlap with events
selected by other \ttbar resonance searches.
Electrons are reconstructed from clusters of EM calorimeter energy deposits matched to
an ID track~\cite{PERF-2017-01} with $|\eta|<2.47$, excluding the barrel and endcap transition region of
$1.37<|\eta|<1.52$, and are calibrated in situ using $Z \rightarrow ee$ decays~\cite{EGAM-2018-01}.
The electron candidates are required to have $\et>25$~\GeV\ and to
satisfy the `tight' identification criteria defined in Ref.~\cite{ATLAS-CONF-2016-024}.
 
Muons are reconstructed by matching tracks reconstructed in the ID and the muon spectrometer.
The muon candidates are required to have $\pt>25$~\GeV\ and $|\eta|<2.5$, and
satisfy the `medium' quality requirements. The selected muons are calibrated in situ using $Z \rightarrow \mu\mu$
and $J/\psi \rightarrow \mu\mu$ decays~\cite{PERF-2015-10}.
 
Similarly to large-$R$ jets, small-$R$ jets are also built from three-dimensional topological clusters of
energy deposits in the calorimeter, but reconstructed with a radius parameter of $R=0.4$.
Small-$R$ jets from pile-up interactions are suppressed by applying the jet vertex tagger~\cite{PERF-2014-03},
which uses information from tracks associated with the hard-scatter and pile-up vertices.
Electron and muon candidate tracks are required to be associated with the primary vertex using criteria based
on the longitudinal and transverse impact parameters.
To avoid the misidentification of jets as electrons and electrons from heavy-flavour hadron decays,
the closest small-$R$ jet within a cone of size $\Delta R_y = \sqrt{(\Delta y)^{2} + (\Delta\phi)^{2}} = 0.2$ around
a reconstructed electron is removed.\footnote{The rapidity is defined as $y= \frac{1}{2} \ln\frac{E+p_z}{E-p_z}$ where $E$ is the energy and $p_z$ is the longitudinal component of the momentum along the beam direction.}
If an electron is then found within $\Delta R_y = 0.4$ of a  small-$R$ jet, the electron
is removed. If a muon is found within $\Delta R_y = 0.04+10\,\text{\GeV}/\pt^{\mu}$ of a  small-$R$ jet (where
$\pt^{\mu}$ is the muon transverse momentum), the muon is removed if the jet contains at least three
tracks, otherwise the jet is removed.
 
\subsection{Event selection and categorization}
\label{subsec:categorization}
The analysis uses events selected by triggers that require at least one large-$R$ jet with
$\pt>360$--460~\GeV, depending on the data-taking period.
The large-$R$ jet with the highest \pt in the event (referred to as the leading jet) is required to
have $\pt>500$~\GeV\ to ensure a nearly 100\% trigger efficiency.
Events are further required to contain at least one more large-$R$ jet with $\pt>350$~\GeV\
to enhance the presence of two jets that can each fully contain the top-quark decay products.
Events containing leptons (electrons or muons) are removed to ensure that there is no overlap with
events selected by other lepton+jets or dilepton analyses.
In addition, events containing small-$R$ jets consistent with originating from detector noise or
calorimeter energy deposits associated with non-collision processes are removed.
The invariant mass, \mJJ, of the two highest-\pt large-$R$ jets is required to be larger than 1.4~\TeV\ to avoid a kinematic
bias caused by the jet \pt requirements.
The two leading jets are required to have a separation in azimuthal angle ($|\Delta\phi|$) larger than 1.6 to ensure a back-to-back topology
of top-quark jets. In addition, the rapidity distance between the two leading jets, $|\Delta y_{\text{JJ}}|$, has to be less than 1.8.
This requirement rejects multijet background events with large $|\Delta y_{\text{JJ}}|$ values, which are
dominated by processes with $t$-channel
gluon exchanges. Events selected after applying these cuts (`preselection') are further
refined by the signal region (SR) selections described below.
 
\begin{figure}[t]
\centering
\subfloat[]{
\includegraphics[width=0.45\textwidth]{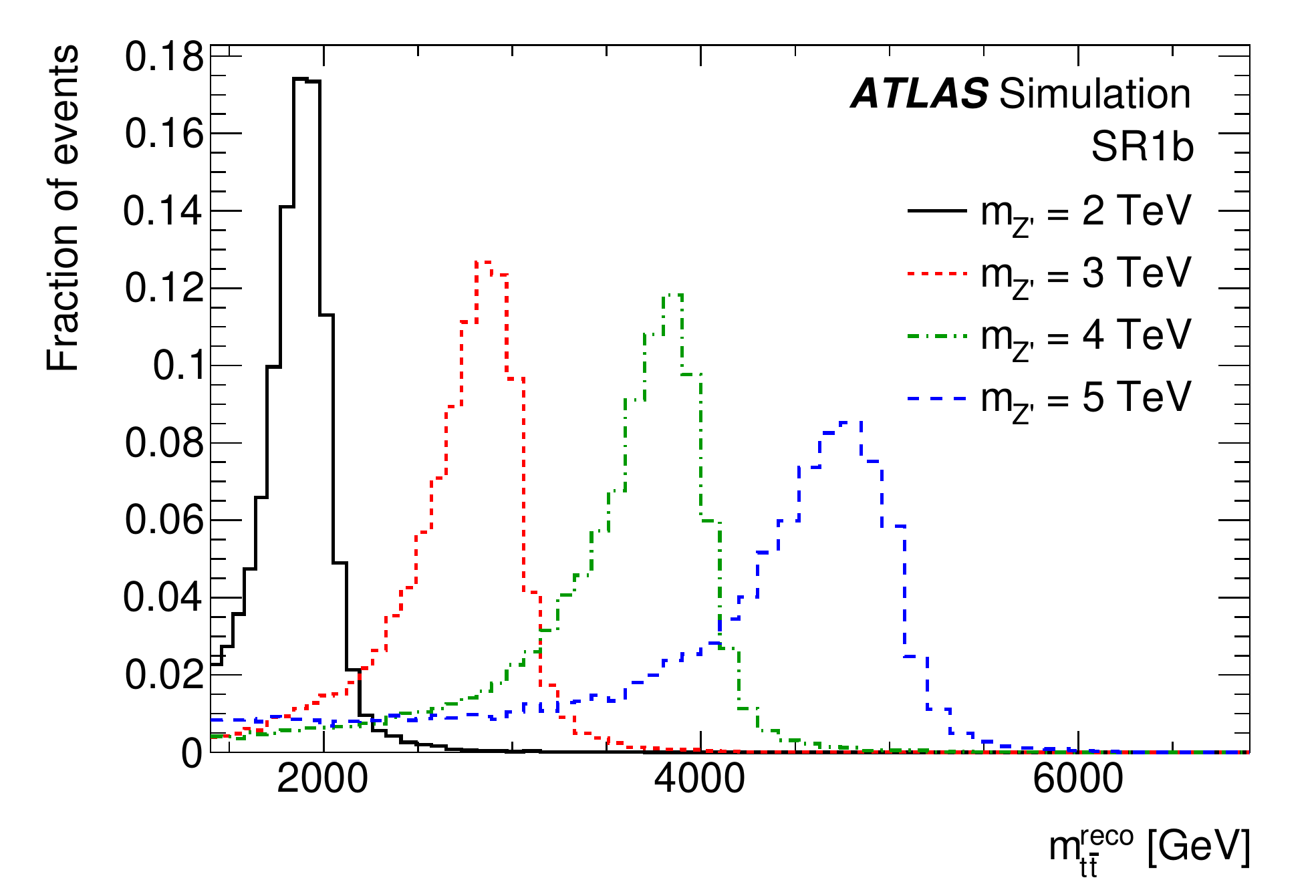}
}
\subfloat[]{
\includegraphics[width=0.45\textwidth]{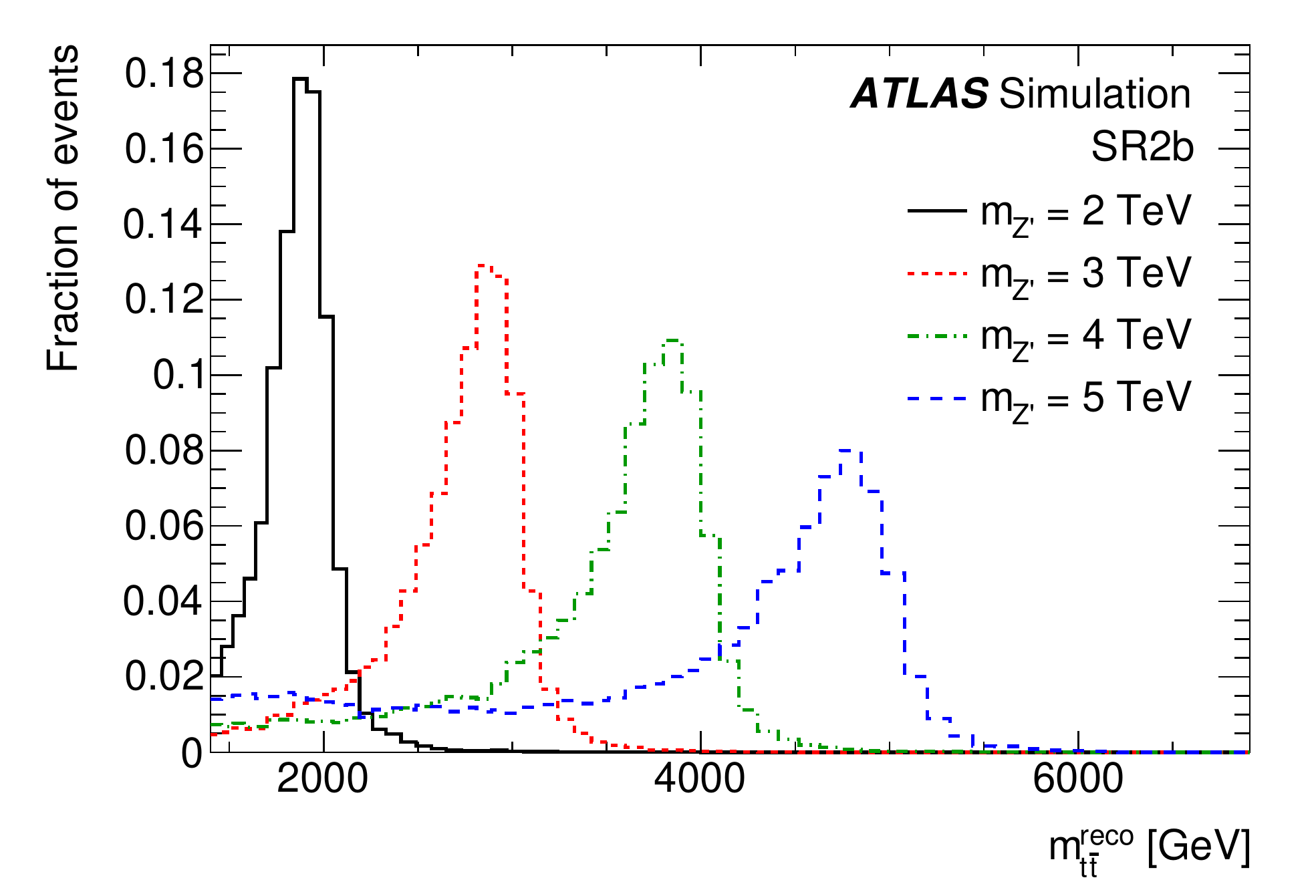}
}
\caption{Normalized \mttreco distributions for simulated $\zprime \to \ttbar$ signal events for \zprime masses of 2, 3, 4 and 5~\TeV\
for (a) SR$1b$ and (b) SR$2b$.}
\label{fig:zprime_mttReco}
\end{figure}

\begin{figure}[t]
\centering
\subfloat[]{
\includegraphics[width=0.45\textwidth]{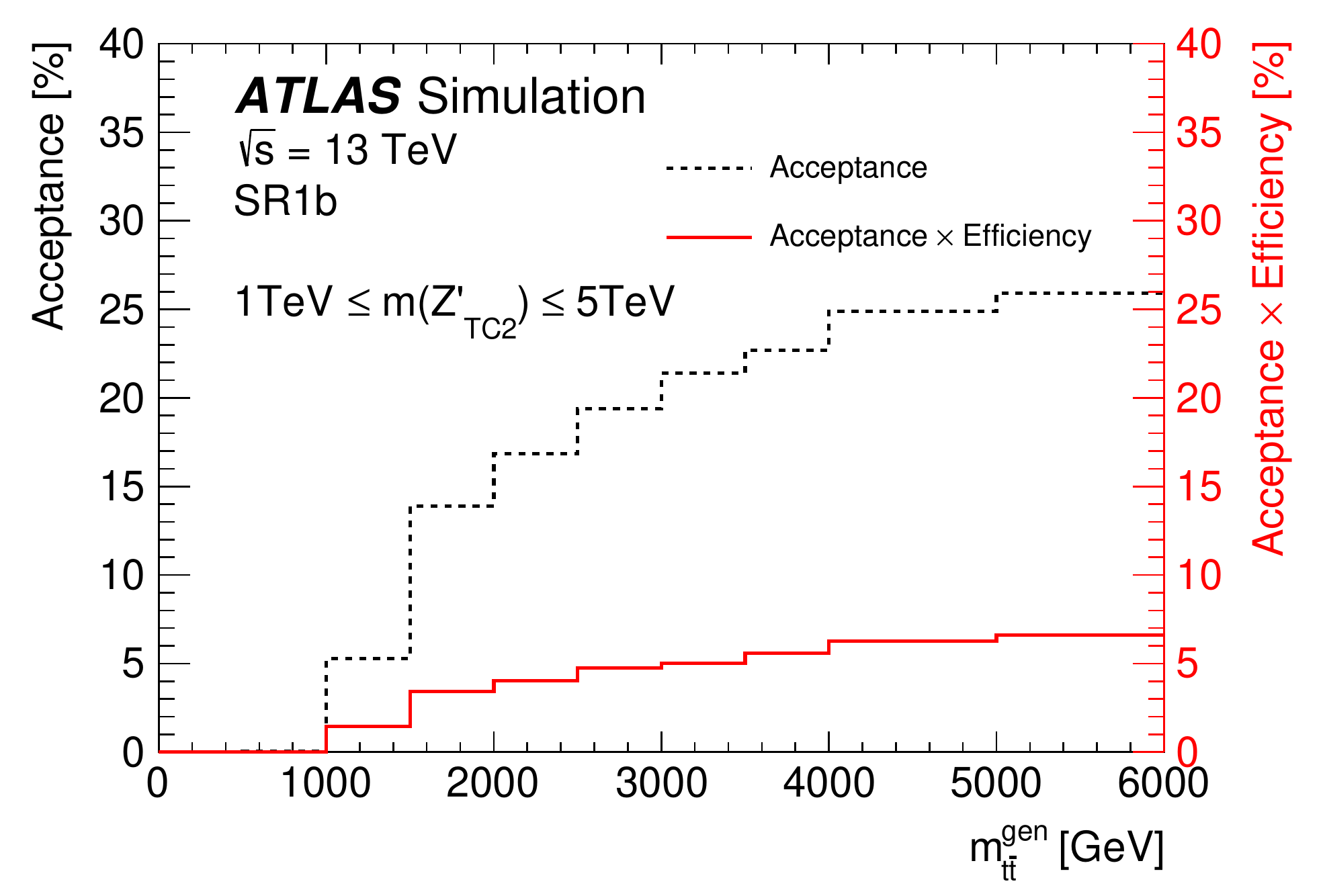}
}
\subfloat[]{
\includegraphics[width=0.45\textwidth]{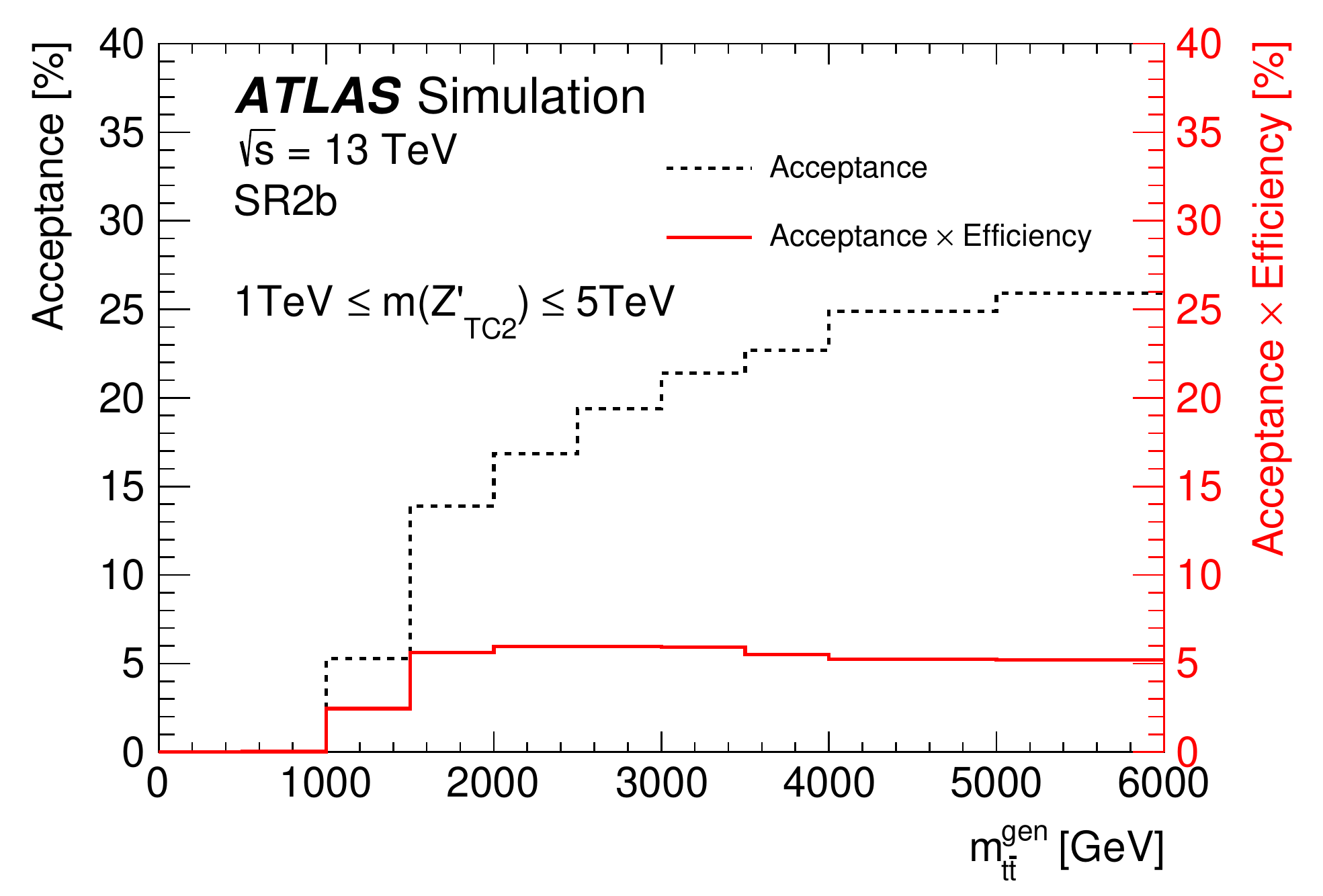}
}
\caption{Acceptance (dashed histograms)  and acceptance times selection efficiency (solid histograms)
as a function of \mttgen in (a) SR$1b$ and (b) SR$2b$.
The acceptance is measured as the fraction of events with two leading truth-contained large-$R$ jets, both satisfying the
kinematic requirements, but not containing generator-level electrons or muons,
as described in Section~\ref{subsec:categorization}.
The acceptance $\times$ efficiency is calculated with respect to the full analysis selections including top- and $b$-tagging
requirements on the two leading large-$R$ jets. The \mttgen is calculated from the momenta of top and anti-top quarks
at the generator level before final-state radiation.
The branching fractions of the \ttbar into all possible final states are included in the acceptance calculation.}
\label{fig:zprime_accXeff}
\end{figure}

In the analysis, the two leading large-$R$ jets in the preselected events are required to be top-tagged. The selected events are further
categorized into two orthogonal SRs by the number of $b$-tagged VR track-jets (\nb) associated with the top-tagged large-$R$ jets
by having a separation $\Delta R<1.0$.
Events satisfying $\nb=1$, i.e.\ exactly one of the two top-tagged jets is associated with a $b$-tagged jet,
are categorized into `SR$1b$'. Events satisfying $\nb=2$, i.e.\ each of the two top-tagged jets is associated with a $b$-tagged jet,
are categorized into `SR$2b$'.
The normalized distributions of the reconstructed \mtt (\mttreco) in the $1b$ and $2b$ signal events are shown in
Figure~\ref{fig:zprime_mttReco} for different masses of the \zprime boson.
The \mttreco distribution has a tail towards masses lower than the generated mass due to
off-shell production of the signal and the growth of low-$x$ partons in the proton PDF
(where $x$ stands for Bjorken-$x$ of the proton).
The acceptance times efficiency as a function of the invariant mass of a top-quark pair at the generator level, \mttgen, is
shown separately for SR$1b$ and SR$2b$ in Figure~\ref{fig:zprime_accXeff}.
The acceptance is measured as the fraction of events with two leading truth-contained large-$R$ jets, both satisfying
the kinematical requirements on the \pt, $\eta$, $\Delta\phi$ and $\Delta y$ described above, but not containing
any generator-level leptons that satisfy certain kinematic selections.\footnote{The generator-level leptons considered in the acceptance calculation are electrons or muons with $\pt>25$~\GeV\ and $|\eta|<2.5$.}
The acceptance $\times$ efficiency is obtained with respect to the full analysis selections including top- and $b$-tagging
requirements on the two leading large-$R$ jets.
The acceptance increases with increasing \mttgen, largely due to the truth-contained requirements.
The acceptance $\times$ efficiencies have different  \mttgen dependence for the two signal regions, mainly caused by the
different $b$-tagging requirements.

\section{Background estimation}
\label{sec:background}
The main backgrounds after applying the selection criteria described in Section~\ref{sec:analysis} are expected to arise
from SM production of \ttbar pairs and multijet events. The background \mttreco distribution in the signal regions
is estimated directly from data by performing a fit with a smoothly falling spectrum.
The appropriate functional form of the spectrum is determined using
combinations of data and simulated events, as described below.
 
\subsection{Background modelling using data and simulation}
\label{subsec:bg_modeling}
In order to ensure the correct functional modelling of the background, the functional form is determined
using expected background spectra in the signal regions. These are obtained by summing the expected distributions of
the SM \ttbar and multijet events.
 
For the \ttbar background, the simulation samples described in Section~\ref{sec:data_sim} are used.
For the multijet background, in addition to the multijet simulation samples in Section~\ref{sec:data_sim},
dedicated control regions (CRs) in data are used to extract the expected \mttreco distributions at low masses.
The distributions extracted from the data are combined with those from the simulation sample,
such that the combined multijet sample provides more events than expected for the data at high masses.
Table~\ref{tab:CR_multijet} shows the CRs defined in data
according to whether the leading and subleading large-$R$ jets pass or fail the top- and/or $b$-tagging requirements.
The SRs defined in Section~\ref{subsec:categorization}
correspond to the regions SR$1b_{\tb\tnb}$, SR$1b_{\tnb\tb}$ and SR$2b_{\tb\tb}$.
The first (last) two subscripts correspond to the leading (subleading) large-$R$ jet, and they indicate that
the jet is tagged ($k$) or not ($\cancel{k}$) where $k=t~(b)$ corresponding to top-tagging ($b$-tagging).
The SR$1b$ signal region consists of the sum of the two regions, SR$1b_{\tb\tnb}$ and SR$1b_{\tnb\tb}$.
The `template region' $\text{TR}_{\tnb\tnb}$ with two top-tags and zero $b$-tags
is dominated by multijet events, with a purity of about 97\%, and hence it is used to extract the shape of
the multijet distribution. The multijet background contributions expected in the SR$i$
($i$ = $1b_{\tb\tnb}$, $1b_{\tnb\tb}$, $2b_{\tb\tb}$),
$N_{\text{exp}}(\text{SR}i)$, is obtained by multiplying the
observed events in the TR, $N(\text{TR}_{\tnb\tnb})$, by scale factors $R$ for
either one or both leading top-tagged jets to be $b$-tagged:
\begin{equation}
N_{\text{exp}}(\text{SR}i) =  N(\text{TR}_{\tnb\tnb}) \times R(\text{TR}_{\tnb\tnb} \to \text{SR}i),
\label{eq:DD1}
\end{equation}
where
\begin{eqnarray}
R(\text{TR}_{\tnb\tnb} \to \text{SR}1b_{\tnb\tb}) &=& \kappa_{t_1b_2} \times \frac{N(\text{A}_{\ntnb\tb})}{N(\text{B}_{\ntnb\tnb})}, \\
R(\text{TR}_{\tnb\tnb} \to \text{SR}1b_{\tb\tnb}) &=& \kappa_{t_2b_1} \times \frac{N(\text{I}_{\tb\ntnb})}{N(\text{H}_{\tnb\ntnb})}
\end{eqnarray}
and
\begin{eqnarray}
R_1(\text{TR}_{\tnb\tnb} \to \text{SR}2b_{\tb\tb}) &=& R(\text{TR}_{\tnb\tnb} \to \text{SR}1b_{\tnb\tb})  \times \kappa_{t_1b_1} \times \frac{N(\text{E}_{\ntb\tnb})}{N(\text{B}_{\ntnb\tnb})}, \\
R_2(\text{TR}_{\tnb\tnb} \to \text{SR}2b_{\tb\tb}) &=& R(\text{TR}_{\tnb\tnb} \to \text{SR}1b_{\tb\tnb})  \times  \kappa_{t_2b_2} \times \frac{N(\text{G}_{\tnb\ntb})}{N(\text{H}_{\tnb\ntnb})}.
\end{eqnarray}
Here the $N$ without subscript represents the observed number of events in the respective region.
 
The scale factors from TR$_{\tnb\tnb}$ to SR$2b_{\tb\tb}$ have two values, $R_1$ and $R_2$, corresponding to the two possible paths:
$\text{TR}_{\tnb\tnb} \to \text{SR}1b_{\tnb\tb} \to \text{SR}2b_{\tb\tb}$ and $\text{TR}_{\tnb\tnb} \to \text{SR}1b_{\tb\tnb} \to \text{SR}2b_{\tb\tb}$, respectively.
The two resulting $N_{\text{exp}}(\text{SR}2b_{\tb\tb})$ values agree within 10\% with the average of the two,
and the average is used as the final prediction for the region SR$2b_{\tb\tb}$.
The $\kappa_{t_1b_2}$ ($\kappa_{t_2b_1}$) factor accounts for the correlation between the top-tagging of the leading (subleading) large-$R$ jet and the $b$-tagging of the subleading (leading) large-$R$ jet and is obtained as follows:
\begin{equation}
\kappa_{t_1b_2} = \left( \frac{N(\text{G}_{\tnb\ntb})}{N(\text{H}_{\tnb\ntnb})} \right) \bigg/ \left( \frac{N(\text{C}_{\ntnb\ntb})}{N(\text{D}_{\ntnb\ntnb})} \right), \hspace{0.75cm}
\kappa_{t_2b_1} = \left( \frac{N(\text{E}_{\ntb\tnb})}{N(\text{B}_{\ntnb\tnb})} \right) \bigg/ \left( \frac{N(\text{F}_{\ntb\ntnb})}{N(\text{D}_{\ntnb\ntnb})} \right).
\end{equation}
Similarly, the $\kappa_{t_1b_1}$ ($\kappa_{t_2b_2}$) factor accounts for the correlation between the top-tagging of the leading (subleading) large-$R$ jet and the $b$-tagging of the leading (subleading) large-$R$ jet, as follows:
\begin{equation}
\kappa_{t_1b_1} = \left( \frac{N(\text{I}_{\tb\ntnb})}{N(\text{H}_{\tnb\ntnb})} \right) \bigg/ \left( \frac{N(\text{F}_{\ntb\ntnb})}{N(\text{D}_{\ntnb\ntnb})} \right), \hspace{1.0cm}
\kappa_{t_2b_2} = \left( \frac{N(\text{A}_{\ntnb\tb})}{N(\text{B}_{\ntnb\tnb})} \right) \bigg/ \left( \frac{N(\text{C}_{\ntnb\ntb})}{N(\text{D}_{\ntnb\ntnb})} \right).
\label{eq:DD5}
\end{equation}
 
These calculations in Eqs.~(\ref{eq:DD1})--(\ref{eq:DD5}) are performed bin-by-bin for the \mttreco distributions in all relevant CRs.
The average values of the $\kappa_{t_1b_2}$ and $\kappa_{t_2b_1}$ factors
range between 0.94 and 1.0 while those of $\kappa_{t_1b_1}$ and
$\kappa_{t_2b_2}$ have larger values around 1.6--1.7 in the \mttreco region of 1.4--6.0~\TeV.
The large difference between the $\kappa$ factors indicates that the variation of multijet event yields
due to the top($b$)-tagging depends on whether the same top($b$)-tagged jet in the event is also $b$(top)-tagged or not.
 
The resulting multijet \mttreco distributions for the SRs are combined with the simulated sample of \PYTHIA multijet events
to reduce statistical fluctuation at high mass. This is carried out by using the simulated \PYTHIA sample instead of
the data-derived sample in the \mttreco region where the former has higher statistical power than the latter.
It turns out that the simulated sample has different \mttreco shape than the data-derived sample and the difference is
quantified as a linear fit to the ratio of data-derived to simulated event yields in \mttreco.
Therefore, the \mttreco shapes of the simulated sample are corrected in the SRs by applying the fit values in bins of
the \mttreco distributions.
In order to avoid a dependence on the choice of \mttreco value
where the data-derived and simulated multijet samples are combined,
the switching point is varied in steps of 140--200~\GeV\ between 2260 and
3030~\GeV\ when the background modelling uncertainty is considered (Section~\ref{subsec:background_modeling_uncertainties}).

\begin{table}[!htbp]
\begin{center}
\caption{Event categorization used to model the multijet background from data according to whether
the leading and subleading large-$R$ jets are top-tagged or $b$-tagged.
If the large-$R$ jet is top-tagged, it is denoted by $t$, and otherwise by \cancel{$t$}, as indicated in the left column or
in the bottom row. Similarly, if the large-$R$ jet is $b$-tagged, it is denoted by $b$, and otherwise by \cancel{$b$}.
The percentages in parentheses show the expected fractions of SM \ttbar events obtained using the \ttbar and multijet
simulation samples. Non-\ttbar or non-multijet background events are negligible.
The signal regions, SR$1b$ and SR$2b$, are coloured in red,
the template region (TR) in grey and the rest of the control regions A-I in light blue.
}
\vspace{0.4cm}
\small{
\newcommand\T{\rule{0pt}{2.6ex}}
\definecolor{lightblue}{rgb}{0.67,0.8,0.95}
\definecolor{lred}{rgb}{0.95,0.5,0.5}
\definecolor{cadet}{rgb}{0.57,0.64,0.69}
\begin{tabular}{c|c|c|c|c|c|}
\hhline{|~|-----|}
\multirow{4}{*}{\rotatebox{90}{\parbox{2cm}{Subleading large-$R$ jet}}}& $tb$ &
\cellcolor{lightblue}A (6.1\%) & &
\cellcolor{lred}SR$1b$ (23\%) & \cellcolor{lred}SR$2b$ (90\%) \T \\ \hhline{|~|-----|}
&  $t\cancel{b}$ & \cellcolor{lightblue}B (0.5\%) & \cellcolor{lightblue}E (1.8\%) &
\cellcolor{cadet}TR (2.6\%) & \cellcolor{lred}SR$1b$ (28\%) \T \\ \hhline{|~|-----|}
&  $\cancel{t}b$ & \cellcolor{lightblue}C (0.4\%) &  &
\cellcolor{lightblue}G (2.3\%) & \T \\ \hhline{|~|-----|}
& $\cancel{t}\cancel{b}$ & \cellcolor{lightblue}D ($<0.1$\%) & \cellcolor{lightblue}F (0.3\%) &
\cellcolor{lightblue}H (0.4\%) & \cellcolor{lightblue}I (6.7\%) \T \\ \hhline{|~|-----|}
& & $\cancel{t}\cancel{b}$ & $\cancel{t}b$ & $t\cancel{b}$ & $tb$ \T \\ \hhline{|~|-----|}
\multicolumn{1}{c}{}& \multicolumn{5}{c}{Leading large-$R$ jet} \T \\
\end{tabular}
}
\label{tab:CR_multijet}
\end{center}
\end{table}
 
\subsection{Determination of background parameterization}
\label{subsec:bg_parameterization}
The expected background \mttreco distributions obtained in Section~\ref{subsec:bg_modeling} are used,
along with the simulated SM \ttbar samples, to determine
the functional forms of the fits to the SR data. To do this, a set of 1000 background distributions
(referred to as $S_{\text{bkg}}$ hereafter) is created for
each SR by bin-wise variation of the nominal distribution assuming Poisson fluctuations.
A typical bin width is chosen using the resolution of the reconstructed \mttreco distribution and is 60~\GeV\ at $\mttreco=1.4$~\TeV, increasing to 100 (130)~\GeV\ at $\mttreco=4$ (6)~\TeV.
For each bin of the background \mttreco distribution, the square-root of the bin content is assigned as the uncertainty.
 
The following parameterized form~\cite{EXOT-2016-33} is used for the fit function:
\begin{equation}
F(x) = p_0(1 - x)^{p_1}x^{p_2+p_3\log(x)+p_4\log(x)^2},
\label{eq:fit_function}
\end{equation}
where $x=\mttreco/\sqrt{s}$, $p_0$ is a normalization factor and $p_1$ through $p_4$ are free parameters controlling the
shape of the \mttreco distributions. The function can be used as a two- or three-parameter function by setting $p_3=p_4=0$ or $p_4=0$, respectively. The number of shape parameters is then determined by performing
fits with individual functions to the $S_{\text{bkg}}$ set of background distributions and evaluating
the goodness of the individual fits.
The evaluation of the goodness of the fit is performed
using $\chi^2$ and BumpHunter~\cite{Choudalakis:2011qn} test statistics.
The BumpHunter is a hypothesis-testing tool to look for local excesses or deficits in data relative to the background
(obtained from fits in this analysis).
After confirming that all the two-, three- and four-parameter functions are qualified, a Wilks' test~\cite{wilks1938} is repeated over the $S_{\text{bkg}}$ set of background distributions to determine how many parameters are needed to describe them.
Finally, the functional form with three shape parameters ($p_4=0$) is selected for both SR$1b$ and SR$2b$.
The fit with the three-shape-parameter function is performed to the data to estimate background in each of the SRs.
 
\section{Systematic uncertainties}
For uncertainties in the signal prediction, experimental uncertainties associated with the reconstruction and
calibration techniques are considered in the analysis. In addition, an uncertainty associated with the interpolated signal
templates and the statistical uncertainty of the simulated samples are taken into account as detailed below.
Since the background is directly estimated from data using a functional fit, only the uncertainties
associated with the fit method and the statistical uncertainty of the data are taken into account for the background modelling.
Each source of experimental systematic uncertainty in the signal is treated as fully correlated across signal regions.
The background uncertainties are considered to be uncorrelated as they are estimated from statistically independent
samples.
 
\subsection{Signal modelling uncertainties}
The experimental systematic uncertainties considered for the signal prediction are dominated by uncertainties associated with
the large-$R$ jet energy scales (JES), jet mass scale (JMS), jet energy resolutions (JER), jet mass resolution (JMR), and the top-tagging and $b$-tagging efficiencies.
 
The uncertainty in the scale of large-$R$ jet \pt and mass is evaluated by
comparing the ratio of calorimeter-based to track-based measurements in multijet data and simulation~\cite{PERF-2015-04,ATLAS-CONF-2016-035}.
The uncertainty in the jet \pt resolution~\cite{PERF-2011-04} is obtained by smearing the MC jet \pt using a Gaussian function
with a relative width of 0.02 and calculating the resulting change from the unsmeared distribution.
This is obtained as a one-sided systematic uncertainty and then symmetrized in the rest of the analysis.
The jet mass resolution~\cite{ATLAS-CONF-2016-035} uncertainty, derived specifically in the context of top-quark-initiated jets,
is obtained by smearing the MC jet mass using a Gaussian function with a width corresponding to a 20\% relative uncertainty
on the jet mass resolution.
 
The top-tagging efficiency for hadronically decaying top quarks is measured using both data and simulation
samples enriched in \ttbar events with one-lepton final states~\cite{JETM-2018-03}. The difference between the data and
simulation efficiencies is taken as a correction factor to the simulated signal yield. The uncertainty in
the correction factor, estimated to be 10--15\% (per jet) depending on the top-quark jet \pt, is propagated through the signal yields in the SRs.
An additional top-tagging systematic uncertainty is considered at high \pt, beyond a top-quark jet \pt of approximately 1~\TeV,
where the correction factor is not measured due to the limited number of one-lepton \ttbar events in data.
This uncertainty is evaluated as the variation of the efficiencies from simulated \ttbar samples with different conditions applied to the $\GEANT4$ calorimeter shower model and the detector material,
and is 1--6\% for the top-quark jet \pt between 1 and 3~\TeV.
This additional uncertainty is added in quadrature to the correction factor uncertainty at the highest available \pt bin
to obtain the top-tagging systematic uncertainty beyond that \pt bin.
The components of jet \pt~scale and top-tagging uncertainties associated with the same sources of systematic
uncertainties are varied together in the statistical analysis procedure.

The uncertainty in the $b$-tagging efficiency for $b$-quark-induced jets is accounted for in a way similar to
that for the top-tagging uncertainty.
The $b$-tagging efficiencies in data and simulated events are compared
using \ttbar events with final states containing two leptons, and the correction factor is derived as the difference
between data and simulation efficiencies~\cite{PERF-2016-05}.
The $b$-tagging uncertainties in the correction factor are assessed in various kinematic regions,
separately for $b$-jets, $c$-jets, and light-flavour jets.
An additional $b$-tagging uncertainty at high \pt beyond the measured \pt range is also taken into account by
an extrapolation technique.
The uncertainties are then decomposed into various independent components for each of the flavour categories.
 
The uncertainty in the combined 2015--2018 integrated luminosity is 1.7\% \cite{ATLAS-CONF-2019-021}, obtained using the LUCID-2 detector \cite{LUCID2} for the primary luminosity measurements.
The pile-up modelling uncertainty including the uncertainty associated with the pile-up suppression by the jet vertex tagger
is considered for simulated events. The uncertainties associated with
lepton reconstruction and identification are negligible compared with other uncertainties.
The uncertainty associated with the interpolated signal is taken into account by comparing expected limits obtained
using the simulated and interpolated signal samples at the same masses.
The uncertainty is estimated to be at most 5\%, and therefore a 5\% uncertainty is assigned to the signal event yield,
along with the statistical uncertainty due to the limited number of simulated events.
 
A summary of the post-fit impacts on the 2 and 4~\TeV\ \zprime signal yields is given in Table~\ref{tab:postfitSyst}.
They are expressed as percentage deviations from the nominal yields
due to the large-$R$ jet related uncertainties, top- and $b$-tagging uncertainties as well as the luminosity and
pile-up modelling uncertainties.
The impacts on the signal yields are amplified from the per-jet \pt-scale and top-tagging uncertainties
due to the presence of two top-quark jets in the events.
The impact from the $b$-tagging uncertainty is larger in the SR$1b$ than the SR$2b$ at the same signal masses.
This is understood as follows: signal events in both SR$2b$ and, to a lesser extent, SR$1b$ are dominated
by events with two $b$-quarks, each associated with one of the two leading large-$R$ jets.
Both jets (exactly one jet) are required to be $b$-tagged in the SR$2b$ (SR$1b$).
This results in an increased impact from the $b$-tagging uncertainty for the SR$1b$ because,
for the $b$-tagging criteria used in this analysis, the uncertainty in the inefficiency correction factor for
the non-$b$-tagged jet is about a factor of 3 larger than that in the efficiency correction factor for the $b$-tagged jet.
The impact from the $b$-tagging uncertainty further increases for signals with mass of 4~\TeV\ in the SR$1b$
due to the additional contribution from the extrapolation to high \pt.

\begin{table}
\caption{Summary of the post-fit impact of systematic uncertainties on the expected numbers of $Z^{\prime}_{\text{TC2}}$ signal events
(expressed as percentage deviations from the nominal predictions)
for masses of 2 and 4~\TeV\ in each of the signal regions.
The largest absolute value due to the $\pm$1$\sigma$ variations is shown for each uncertainty source.
The components of top-tagging uncertainties associated with the JES uncertainties are included in the numbers at the row
labeled JES.}
\label{tab:postfitSyst}
\centering
\begin{tabular}{l | l | l | l | l l}
\toprule
\textbf{Source} & \multicolumn{2}{c |}{\textbf{2 \TeV\ $Z^{\prime}$ [\%]}} & \multicolumn{2}{c}{\textbf{4 \TeV\ $Z^{\prime}$ [\%]}}\\
\midrule
& SR$1b$ & SR$2b$ & SR$1b$ & SR$2b$ \\
\midrule
JES  & 35& 34& 47& ~~44 \\
JMS  & ~~5.0& ~~4.3& ~~9.5& ~~~~7.9\\
JER  & ~~0.1& ~~0.1& ~~0.1& $<0.1$\\
JMR  & ~~3.9& ~~4.0& ~~8.0& ~~~~8.0\\
$b$-tagging  & 14& ~~5.0& 23& ~~~~5.3\\
Top-tagging  & ~~9.0& ~~9.3& 10& ~~10\\
Luminosity & ~~1.7 & ~~1.7 & ~~1.7 & ~~~~1.7\\
Pile-up modelling & ~~0.2 & ~~0.2 & ~~0.2 & ~~~~0.2\\
\bottomrule
\end{tabular}
\end{table}

\subsection{Background modelling uncertainties}
\label{subsec:background_modeling_uncertainties}
The uncertainty in the background estimate is primarily associated with the choice of functional form and the fit range
used to estimate the background. With a given fit function with a certain number of shape parameters and fit range,
there is an ambiguity in the fit result due to intrinsic limitations associated with the chosen fit.
Because of this, when a fit is performed to the background-only distribution,
the fit creates a systematic difference between the estimated and real backgrounds, producing a
`spurious' signal.
This systematic difference, referred to as the spurious-signal uncertainty,
is evaluated in the statistical analysis as a bias in the signal estimate obtained from a signal-plus-background fit
to the \mttreco distribution constructed under the background-only hypothesis.
The signal model used in the signal-plus-background fit is extracted from the simulated \zprime samples
by performing fits to the reconstructed \mttreco spectra
with the sum of a Gaussian function and a Crystal Ball function.
The background model is the three-shape-parameter function of Eq.~(\ref{eq:fit_function}).
The spurious-signal uncertainty is obtained from a signal-plus-background fit performed over the $S_{\text{bkg}}$
set of background distributions used to determine the fit functions in Section~\ref{subsec:bg_parameterization}.
In the fit the switching point for the data-derived and simulated multijet samples is varied in each $S_{\text{bkg}}$ set of
background distributions (Section~\ref{subsec:bg_modeling}).
The spurious signal uncertainty (taken from the average result of the fits) is estimated as a function of the generated signal mass and
quantified relative to the statistical uncertainty of the background prediction within $\pm20$\% around the signal peak.
It decreases monotonically from 80\% to 2\% between 1.75 and 5~\TeV\ for the $1b$ signal region.
For the $2b$ signal region, it is approximately constant within 30--40\% up to 4~\TeV,
then increases  significantly and reaches 200\% at 5~\TeV.
The increase of the spurious-signal uncertainty at masses above 4~\TeV\ in the 2$b$ signal region
is associated with the fact that the background distribution falls more rapidly than that in the $1b$ signal region,
causing the function to be less constrained in the high-mass region and thus producing an increased amount
of spurious signal.
An additional uncertainty associated with the statistical uncertainty of the expected background in the signal regions
is also considered by propagating the impact of the uncertainties in the values of the fit parameters in Eq.~(\ref{eq:fit_function})
to the \mttreco spectra.\footnote{The fit parameter uncertainties are re-evaluated by using another set of background
distributions obtained assuming Poisson uncertainties instead of the uncertainties from the square-root of bin contents. The impact on the fit parameter uncertainties from this change turns out to be negligible.}
 
 
\section{Statistical analysis}
The statistical interpretation of the data consists of two steps: a model-independent test for the presence of
local deviations from smooth \mttreco spectra in data using BumpHunter, and hypothesis testing of data with the benchmark signal model
using the profile likelihood ratio.
 
The BumpHunter test quantifies the significance of local excesses or deficits in the \mttreco distributions,
taking into account the look-elsewhere effect~\cite{Lyons:1900zz, Gross:2010qma}
associated with the scanned mass ranges.
 
The hypothesis testing with the benchmark signal is performed using a binned maximum-likelihood fit
to the \mttreco distributions that is based on the expected signal and background yields.
The likelihood model is defined as:
 
\begin{equation}\nonumber
{\cal L} = \prod_i P_{\text{pois}}(n_i|\lambda_i) \times {\cal N}({\bm \theta})
\end{equation}
 
where $P_{\text{pois}}(n_i|\lambda_i)$ is the Poisson probability to observe $n_i$ events when $\lambda_i$ events are
expected in bin $i$ of the \mttreco distribution, and ${\cal N}({\bm \theta})$ is a series of Gaussian or log-normal distributions for
the nuisance parameters, ${\bm \theta}$, corresponding to the systematic uncertainties related to the signal and background
yields in each bin. The $\lambda_i$ is expressed as
$\lambda_{i} = \mu s_{i} ({\bm \theta}) + b_{i}({\bm \theta})$
with $\mu$ being the signal strength, defined as a signal cross-section in units of the theoretical prediction, determined by the fit,
and $s_{i}( {\bm \theta })$ and $b_{i}( {\bm \theta })$ being the expected numbers of signal and background events, respectively.
Nuisance parameters are allowed to float in the fit and thus vary the normalization and shape of the signal \mttreco distribution as well as the shape of the background \mttreco distribution.
 
The information about $\mu$ is extracted from a likelihood fit to data under the signal-plus-background hypothesis,
using a test statistic based on the profile likelihood ratio. The distributions of the test statistic under the signal-plus-background and background-only hypotheses
are obtained using the asymptotic formulae~\cite{asymptotics}. The systematic uncertainties with the largest post-fit impact on $\mu$ at $m=2$~\TeV\ and $m=4$~\TeV\ are the fit parameter uncertainty and the spurious-signal uncertainty for the background and the JES uncertainties for the signal.
The level of agreement between the observed data and the background prediction is assessed by computing the local $p_0$-value,
defined as the probability to observe an excess at least as large as the
one observed in data, under the background-only hypothesis.
The global $p_0$-value is computed by considering the look-elsewhere effect due to multiple testing on the signal mass points.
Expected and observed upper limits are set at 95\% confidence level (CL) on the cross-section times
branching fraction (\sigmaBR) of new particles decaying into a \ttbar pair
using the CL$_\text{s}$ prescription~\cite{CLs_2002}.
Cross-checks with the test statistic sampled using pseudo-experiments are performed to test the accuracy of the asymptotic
formula in the high-mass region beyond 2~\TeV. The \sigmaBR limits from the asymptotic approximation are found to be stronger than
those from the pseudo-experiments by at most 20\% at masses above 4~\TeV. The impact on the mass limit from this
approximation is estimated to be below 100~\GeV\ for the \zprime signal considered in this analysis.
 
 
\section{Results}
\begin{figure}[t]
\centering
\subfloat[]{
\includegraphics[width=0.45\textwidth]{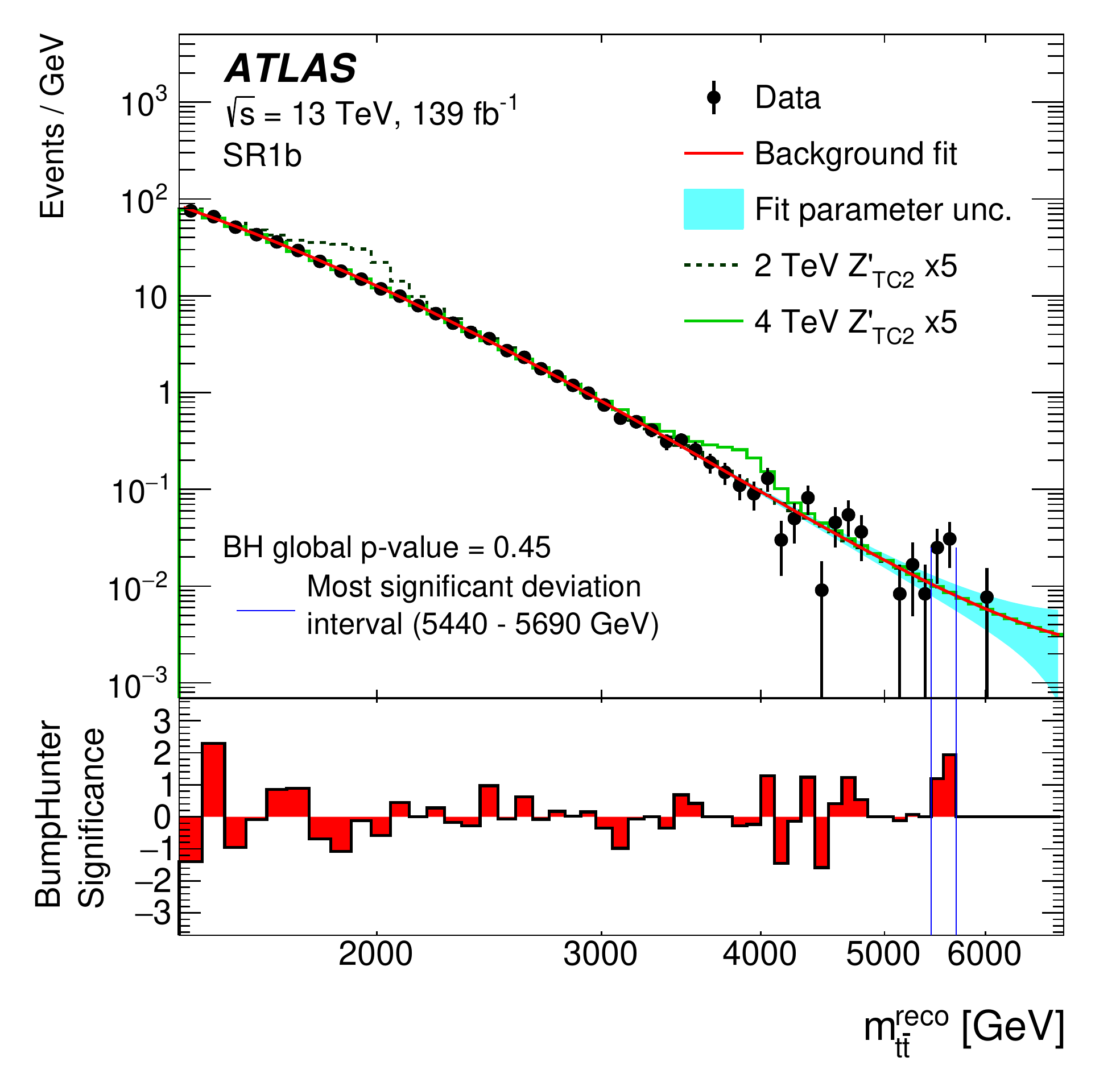}
}
\subfloat[]{
\includegraphics[width=0.45\textwidth]{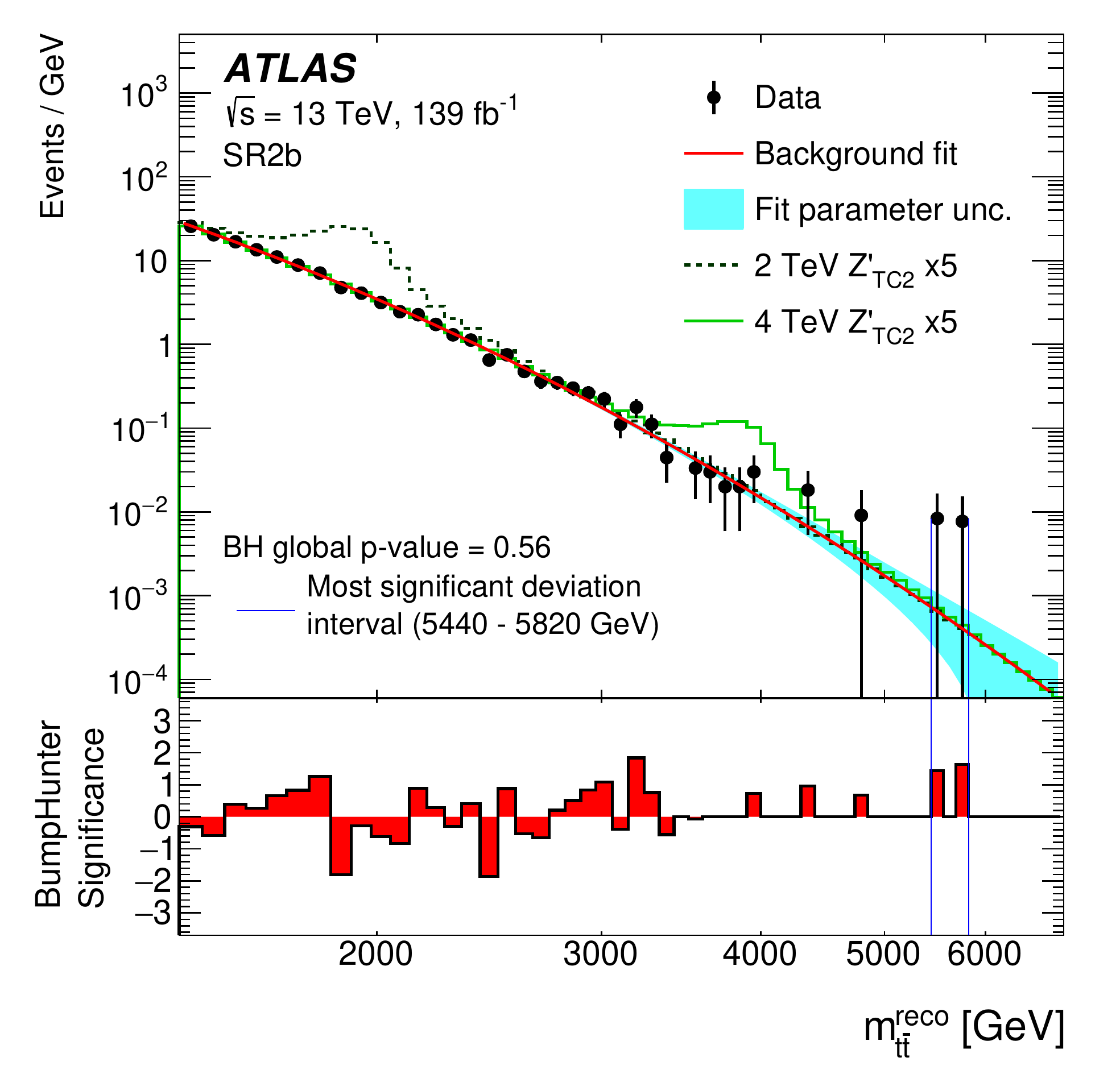}
}
\caption{Observed \mttreco distributions in data for (a) SR$1b$ and (b) SR$2b$,
shown together with the result of the fit with the three-shape-parameter function.
The shaded bands around the fits indicate the effect of the fit parameter uncertainty on the background prediction.
The bin width of the distributions is chosen to be the same as that used in the background parameterization (Section~\ref{subsec:bg_parameterization}).
The predicted \zprime signal distributions with masses of 2 and 4~\TeV\ (multiplied by a factor of 5) are
superimposed on the background prediction. The lower panel shows the significance of data
with respect to the background prediction from the fit, calculated in continuous mass intervals scanned over
the binned \mttreco distributions.
The two vertical lines extending between the upper and lower panel represent the most significant deviation interval.
The global $p$-value of the interval is 0.45 (0.56) for SR$1b$ (SR$2b$).}
\label{fig:SR_data_fit}
\end{figure}

The observed \mttreco distributions in the two SRs with fits using the three-shape-parameter function are
shown in Figure~\ref{fig:SR_data_fit}. The observed number of data events is 26\,964 (8160) in SR$1b$ (SR$2b$).
The BumpHunter tests for the compatibility of the data and the background prediction show that
the fit describes the data well for both SRs. The interval with the most significant deviation is 5.44--5.69~\TeV\ for
SR$1b$ and 5.44--5.82~\TeV\ for SR$2b$ with the corresponding
global $p$-values of 0.45 and 0.56, respectively.
The parameter values of the fit functions determined from the fits to the data are provided in HEPData~\cite{hepdata}.
The parameter values are consistent with those obtained from fits to the $S_{\text{bkg}}$ set of expected
background distributions, used in Section~\ref{subsec:bg_parameterization}.

\begin{figure}[t]
\centering
\includegraphics[width=0.80\textwidth]{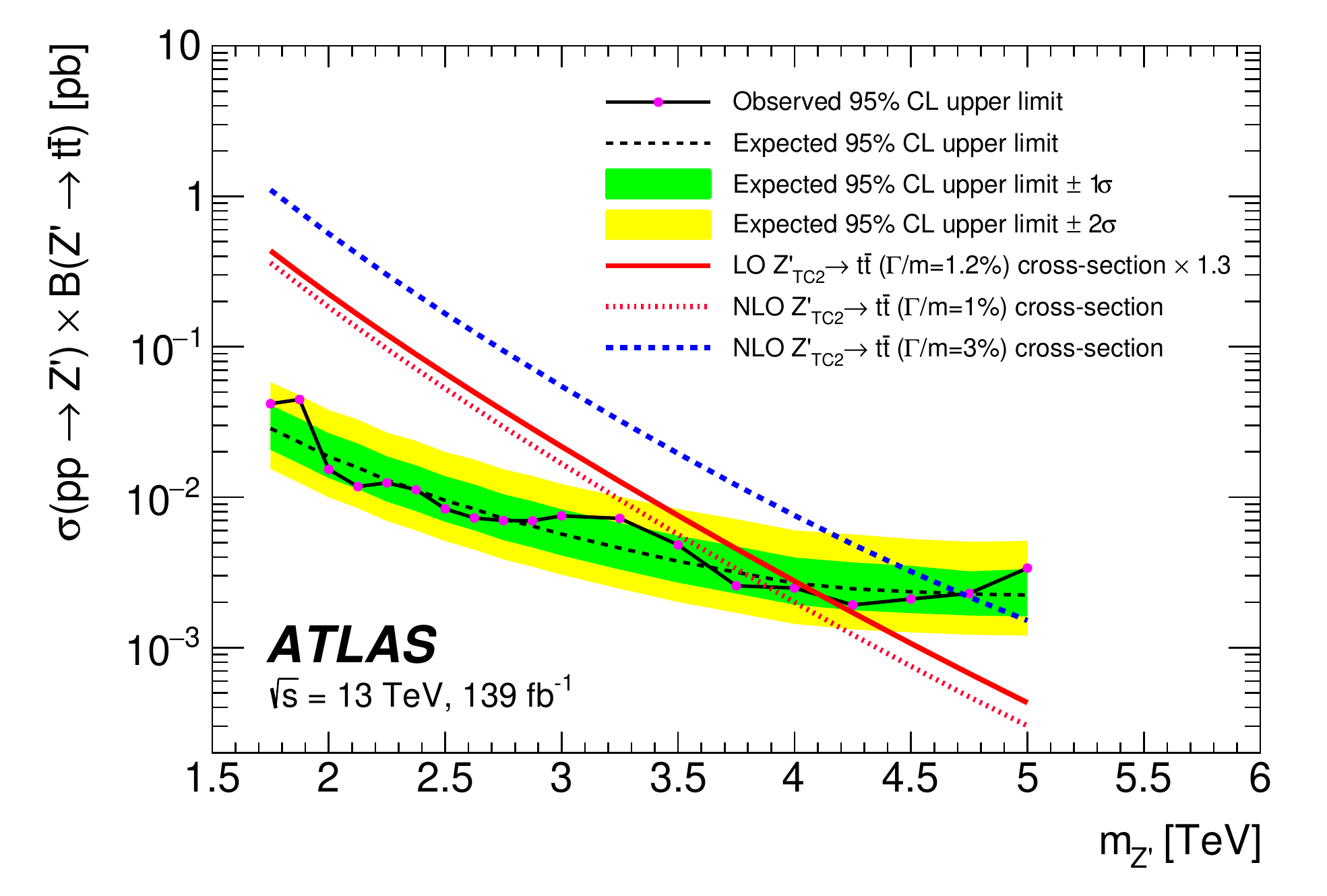}
\caption{Observed and expected upper limits on the cross-section times branching fraction of the $\zprime \to \ttbar$
as a function of the \zprime mass. The NLO theory cross-sections times branching fraction for the \zprime with
$\Gamma/m=1$\% and $3\%$ are shown by the red dotted and blue dashed lines, respectively.
Shown by the red solid line is a NLO prediction, obtained by multiplying the LO theory cross-section times branching fraction
by a factor 1.3~\cite{Bonciani:2015hgv}, for the \zprime with $\Gamma/m=1.2$\%.}
\label{fig:zprime_limit}
\end{figure}

With the \zprime signal used in this analysis, the minimum local $p_0$-value is found to be 0.06 ($1.6\sigma$)
at \zprime mass of 1.88~\TeV\ in the mass range between 1.75 and 5~\TeV.
In the absence of a significant excess above the background prediction, 95\% CL upper limits on
\sigmaBR are calculated at each mass value of the \zprime signal model.
The expected and observed upper limits on the \sigmaBR of $\zprime\to\ttbar$ are presented in
Figure~\ref{fig:zprime_limit}. The results from the two SRs are statistically combined to obtain these limits.
From the comparison with the \sigmaBR at NLO for the \zprime with $\Gamma/m=1\%$
and 3\%, the \zprime masses up to 3.9 and 4.7~\TeV, respectively, are excluded at 95\% CL.
For the \zprime with $\Gamma/m=1.2\%$ and the LO \sigmaBR multiplied by 1.3 (scaled to NLO prediction),
masses up to 4.1~\TeV\ are excluded at 95\% CL.
The upper limits on \sigmaBR are provided only up to 5~\TeV\ for the \zprime signal mass
because of the large spurious-signal uncertainty exceeding 200\%
at masses beyond 5~\TeV, making the limit calculation unreliable at masses larger than $\sim5.2$~\TeV.
The expected sensitivity of the present analysis is limited by the statistical uncertainty of the background prediction
over the full mass range, except at high mass beyond 4.5~\TeV\
where the systematic uncertainty due to the spurious signal dominates the statistical uncertainty.

\FloatBarrier
\section{Conclusion}
This paper presents a search for new massive particles decaying into \ttbar in the fully hadronic final state using
\lumi of $pp$ collision data recorded at $\sqrt{s}=13$~\TeV\ with the ATLAS detector at the LHC.
The search focuses on the mass range above 1.4~\TeV. It uses an improved top-quark tagging based on
a multivariate classification algorithm with a deep neural network and $b$-hadron identification with variable-radius
track-jets for highly boosted top quarks. The background is estimated from a fit to data.
No significant deviation from the background expectation is observed over the search range.
Upper limits are set on the production cross-section times branching fraction for the \zprime boson
in the topcolor-assisted-technicolor model, resulting in the exclusion of \zprime masses up to 3.9 and 4.7~\TeV\
for decay widths of 1\% and 3\%, respectively.
Compared to the previous analysis with the 2015--2016 data set, the improved analysis techniques presented here provide a 65\% improvement in the expected cross-section limit at 4~\TeV\ when using the same data set.
 
\section*{Acknowledgements}

We thank CERN for the very successful operation of the LHC, as well as the
support staff from our institutions without whom ATLAS could not be
operated efficiently.
 
We acknowledge the support of ANPCyT, Argentina; YerPhI, Armenia; ARC, Australia; BMWFW and FWF, Austria; ANAS, Azerbaijan; SSTC, Belarus; CNPq and FAPESP, Brazil; NSERC, NRC and CFI, Canada; CERN; ANID, Chile; CAS, MOST and NSFC, China; COLCIENCIAS, Colombia; MSMT CR, MPO CR and VSC CR, Czech Republic; DNRF and DNSRC, Denmark; IN2P3-CNRS and CEA-DRF/IRFU, France; SRNSFG, Georgia; BMBF, HGF and MPG, Germany; GSRT, Greece; RGC and Hong Kong SAR, China; ISF and Benoziyo Center, Israel; INFN, Italy; MEXT and JSPS, Japan; CNRST, Morocco; NWO, Netherlands; RCN, Norway; MNiSW and NCN, Poland; FCT, Portugal; MNE/IFA, Romania; JINR; MES of Russia and NRC KI, Russian Federation; MESTD, Serbia; MSSR, Slovakia; ARRS and MIZ\v{S}, Slovenia; DST/NRF, South Africa; MICINN, Spain; SRC and Wallenberg Foundation, Sweden; SERI, SNSF and Cantons of Bern and Geneva, Switzerland; MOST, Taiwan; TAEK, Turkey; STFC, United Kingdom; DOE and NSF, United States of America. In addition, individual groups and members have received support from BCKDF, CANARIE, Compute Canada, CRC and IVADO, Canada; Beijing Municipal Science \& Technology Commission, China; COST, ERC, ERDF, Horizon 2020 and Marie Sk{\l}odowska-Curie Actions, European Union; Investissements d'Avenir Labex, Investissements d'Avenir Idex and ANR, France; DFG and AvH Foundation, Germany; Herakleitos, Thales and Aristeia programmes co-financed by EU-ESF and the Greek NSRF, Greece; BSF-NSF and GIF, Israel; La Caixa Banking Foundation, CERCA Programme Generalitat de Catalunya and PROMETEO and GenT Programmes Generalitat Valenciana, Spain; G\"{o}ran Gustafssons Stiftelse, Sweden; The Royal Society and Leverhulme Trust, United Kingdom.
 
The crucial computing support from all WLCG partners is acknowledged gratefully, in particular from CERN, the ATLAS Tier-1 facilities at TRIUMF (Canada), NDGF (Denmark, Norway, Sweden), CC-IN2P3 (France), KIT/GridKA (Germany), INFN-CNAF (Italy), NL-T1 (Netherlands), PIC (Spain), ASGC (Taiwan), RAL (UK) and BNL (USA), the Tier-2 facilities worldwide and large non-WLCG resource providers. Major contributors of computing resources are listed in Ref.~\cite{ATL-SOFT-PUB-2020-001}.
 
 
\printbibliography
 
\clearpage \input{atlas_authlist}

\end{document}

%% file: atlas_authlist.tex
 
\begin{flushleft}
\hypersetup{urlcolor=black}
{\Large The ATLAS Collaboration}

\bigskip

\AtlasOrcid[0000-0002-6665-4934]{G.~Aad}$^\textrm{\scriptsize 102}$,    
\AtlasOrcid[0000-0002-5888-2734]{B.~Abbott}$^\textrm{\scriptsize 128}$,    
\AtlasOrcid[0000-0002-7248-3203]{D.C.~Abbott}$^\textrm{\scriptsize 103}$,    
\AtlasOrcid[0000-0002-2788-3822]{A.~Abed~Abud}$^\textrm{\scriptsize 36}$,    
\AtlasOrcid[0000-0002-1002-1652]{K.~Abeling}$^\textrm{\scriptsize 53}$,    
\AtlasOrcid[0000-0002-2987-4006]{D.K.~Abhayasinghe}$^\textrm{\scriptsize 94}$,    
\AtlasOrcid[0000-0002-8496-9294]{S.H.~Abidi}$^\textrm{\scriptsize 167}$,    
\AtlasOrcid[0000-0002-8279-9324]{O.S.~AbouZeid}$^\textrm{\scriptsize 40}$,    
\AtlasOrcid{N.L.~Abraham}$^\textrm{\scriptsize 156}$,    
\AtlasOrcid[0000-0001-5329-6640]{H.~Abramowicz}$^\textrm{\scriptsize 161}$
% ...
\bigskip
\\

% ...
$^{36}$CERN, Geneva; Switzerland.\\
% ...
$^{40}$Niels Bohr Institute, University of Copenhagen, Copenhagen; Denmark.\\
% ...
$^{53}$II. Physikalisches Institut, Georg-August-Universit\"{a}t G\"ottingen, G\"ottingen; Germany.\\
% ...
$^{94}$Department of Physics, Royal Holloway University of London, Egham; United Kingdom.\\
% ...
$^{102}$CPPM, Aix-Marseille Universit\'e, CNRS/IN2P3, Marseille; France.\\
$^{103}$Department of Physics, University of Massachusetts, Amherst MA; United States of America.\\
% ...
$^{128}$Homer L. Dodge Department of Physics and Astronomy, University of Oklahoma, Norman OK; United States of America.\\
% ...
$^{156}$Department of Physics and Astronomy, University of Sussex, Brighton; United Kingdom.\\
% ...
$^{161}$Raymond and Beverly Sackler School of Physics and Astronomy, Tel Aviv University, Tel Aviv; Israel.\\
% ...
$^{167}$Department of Physics, University of Toronto, Toronto ON; Canada.\\
% ...

\end{flushleft}
